\pdfoutput=1 
\documentclass{article}
\usepackage{amsmath}
\usepackage{graphicx}
\input{tcilatex}
\begin{document}

\title{NON-SMOOTH SPATIO-TEMPORAL COORDINATES IN NONLINEAR DYNAMICS}
\author{V.N. Pilipchuk \\
Wayne State University\\
Detroit, MI 48202}
\maketitle

\begin{abstract}
This paper presents an overview of physical ideas and mathematical methods
for implementing non-smooth and discontinuous substitutions in dynamical
systems. General purpose of such substitutions is to bring the differential
equations of motion to the form, which is convenient for further use of
analytical and numerical methods of analyses. Three different types of
nonsmooth transformations are discussed as follows: positional coordinate
transformation, state variables transformation, and temporal
transformations. Illustrating examples are provided.
\end{abstract}

\section{Introduction}

Discontinuities of states in physical models often represent the result of
intentional idealization of abrupt but still smooth changes in dynamical
characteristics. Such idealizations help to skip complicated details of
modeling on relatively small intervals of the dynamics. From the
mathematical standpoint however, every discontinuity of states breaks the
system and thus increases the system dimension as many as twice even though
the equations may remain the same before and after the discontinuity. This
work will outline and illustrate different ideas of preventing the dimension
increase while dealing with discontinuities of system states. Regardless
physical principals and mathematical specifics, preliminary modification of
descriptive functions represents a common feature of such approaches. Note
that many analytical methods dealing with dynamical systems preliminarily
adapt the equations of motions through different substitutions and
transformations in order to ease further steps of analyses. Although a
universal recipe for such substitutions is rather difficult to suggest,
there are some general principles to follow. For instance, classes of
transformations should comply with classes of systems considered. The term 
\textit{classes} remains intentionally unspecified here since it may
indicate any generic feature of the model, such as linearity or
nonlinearity, a class of smoothness, or other mathematical properties. In
particular, this work focuses on substitutions including nonsmooth or
discontinuous functions.

Generally speaking, differential operations with nonsmooth functions require
generalized interpretations of equalities as integral identities, in other
words, in terms of distributions \cite{Richtmyer:1985Springer}. In linear
cases, such interpretations are usually quite straightforward since
distributions represent linear functionals \cite{Vladimirov:1971Dekker}.
Nonlinear models however impose certain structural constraints on the
presence of non-smooth or discontinuous functions in differential equations 
\cite{Filippov:1988Kluwer}, \cite{Kufner:1980Elsevier}. Moreover, whether or
not some combinations of discontinuous functions are meaningful may depend
upon physical contents of variables participating in such combinations \cite%
{Maslov:1981UMN}. From the mathematical standpoint, physical interpretations
allow for narrowing families of smooth functions that have discontinuities
in their asymptotic limits. As a result, some combinations of discontinuous
functions may acquire certain meanings of distributions. To this end, some
introductory remarks on transitions to nonsmooth or discontinuous limits in
nonlinear cases will be introduced. For that reason, consider, for instance,
the following product of the Heaviside' unit-step function and the Dirac'
delta-function, $\theta (t-1)\delta (t-1)$. Generally speaking, such a
`product' is undefined due to the discontinuity of $\theta (t-1)$. In other
words, no certain number can be prescribed to the integral $\int_{-\infty
}^{\infty }\theta (t-1)\delta (t-1)dt$, unless it is known what kind of
processes both functions actually represented before their transition to
discontinuous limits. For illustrating purposes, let $\delta _{\varepsilon
}(t-1)=\varepsilon ^{-2}\cosh ^{-2}[(t-1)/\varepsilon ^{2}]/2$ be the force
applied to a unit mass whose velocity is zero at $t=-\infty $. Then the
velocity is described by $v(t)=\theta _{\varepsilon }(t-1)\equiv (1+\tanh
[(t-1)/\varepsilon ^{2}])/2$ due to the relationship $\dot{\theta}%
_{\varepsilon }(t-1)=\delta _{\varepsilon }(t-1)$; see Fig. 1 for
illustration.

Based on standard definitions, it can be shown that, in terms of weak
limits, $\delta _{\varepsilon }(t-1)$ gives $\delta (t-1)$ whereas $\theta
_{\varepsilon }(t-1)$ gives $\theta (t-1)$ as $\varepsilon \rightarrow 0$.
Note that the above choice for $\delta _{\varepsilon }(t-1)$ is not unique -
there are many other cases leading to the same result. Regardless magnitude
of $\varepsilon $, however, the work done by this force is calculated as
follows 
\begin{eqnarray*}
\int_{-\infty }^{\infty }\theta _{\varepsilon }(t-1)\delta _{\varepsilon
}(t-1)dt &=& \\
\int_{-\infty }^{\infty }\theta _{\varepsilon }(t-1)\dot{\theta}%
_{\varepsilon }(t-1)dt &=&\frac{1}{2}[\theta _{\varepsilon
}(t-1)]^{2}|_{-\infty }^{\infty }=\frac{1}{2}
\end{eqnarray*}

Therefore, transition to the limit $\varepsilon \rightarrow 0$ gives 
\begin{equation*}
\int_{-\infty }^{\infty }\theta (t-1)\delta (t-1)dt=\frac{1}{2}=\frac{1}{2}%
\int_{-\infty }^{\infty }\delta (t-1)dt
\end{equation*}

In terms of distributions, the above expression means\footnote{%
Strictly speaking, a so-called testing function must be added to the
integrands as a factor in order to completely prove this statement.} 
\begin{equation}
\theta (t-1)\delta (t-1)=\frac{1}{2}\delta (t-1)  \label{theta*delta}
\end{equation}

The left-hand side in (\ref{theta*delta}) can be interpreted as a mechanical
power generated by the impact force $\delta (t-1)$. To some extent, the
result (\ref{theta*delta}) is based on the common sense that such a power
must exist regardless mathematical ambiguity of the left-hand side of
identity (\ref{theta*delta}). Note that interpretation (\ref{theta*delta})
becomes possible due to the `constraint' coupling the two functions, $\dot{%
\theta}_{\varepsilon }(t-1)=\delta _{\varepsilon }(t-1)$, which is supposed
to hold in the limit case $\varepsilon \rightarrow 0$, 
\begin{equation}
\dot{\theta}(t-1)=\delta (t-1)  \label{theta dot}
\end{equation}
were the over dot means Schwartz derivative.

This standard relationship of the theory of distributions as well as its
different variations will be used below. Relationship (\ref{theta*delta})
may also serve as a basis for other definitions, for instance,%
\begin{equation}
\mathtt{sgn}(t-1)\delta (t-1)=0  \label{sgn*delta}
\end{equation}%
where $\mathtt{sgn}(t-1)=2\theta (t-1)-1$.

While using (\ref{theta*delta}) or (\ref{sgn*delta}), both functions $\theta
(t-1)$ and $\delta (t-1)$ must be considered together with their `smooth
prehistories' generated by the same family of smooth functions $\delta
_{\varepsilon }(t-1)$. If the limit functions $\theta (t-1)$ and $\delta
(t-1)$ are based on different families of smooth functions then
relationships (\ref{theta*delta}) or (\ref{sgn*delta}) may not hold; see the
next section for more details.

To conclude this, as follows from the above remarks, despite of the
universal notations, discontinuous functions may still inherit some features
of the generating families of smooth functions. Therefore, analytical
manipulations with discontinuous and delta-functions must account for both
physical content of the problem and mathematical structure of equations. In
the most direct way, such complications can be avoided by considering models
on different time intervals and introducing appropriate matching conditions
for the corresponding pieces of solutions. In many cases, however, the
matching times are a priori unknown and must also be determined from the
same matching conditions. As mentioned at the beginning, this work gives an
overview of another approaches satisfying the matching conditions
automatically through specific non-smooth transformations of variables.
Briefly, the paper is organized as follows: 1) Caratheodori equations and
discontinuous substitutions \cite{Filippov:1988Kluwer} for systems under
external pulses and wave propagation problems, 2) non-smooth coordinate
space transformations for impact systems with elastic perfectly stiff
constraints and possible extensions on non-elastic constraints \cite%
{Zhuravlev:1976MTT}, \cite{Zhuravlev:1978PMM}, \cite{Zhuravlev:1988Nauka},
3) non-smooth state (phase) space transformation \cite{Ivanov:1993PMM}, \cite%
{Ivanov:1994JSV}, \cite{Ivanov:1997International} and applications to
modeling the impact dynamics with an arbitrary coefficient of restitution,
4) non-smooth temporal transformations \cite{Pilipchuk:1985PMM}, \cite%
{Pilipchuk:1988DANUkrUkr} for periodic and non-periodic motions, analytical
and semi-analytical tools, unstable periodic and chaotic motions, 5)
different combined approaches \cite{Thomsen:2008JSV}, \cite%
{Fidlin:2005Springer}, \cite{Pilipchuk:2000JSV}.

\section{Nonsmooth coordinate transformations}

\subsection{Dynamical systems with distributions included as summands:
linear discontinuous transformations}

Different classes of differential equations, linear and nonlinear, with
additively involved distributions were considered Filippov \cite%
{Filippov:1988Kluwer}. In particular, the methods of reducing such equations
to Caratheodori equations were described. Briefly, such methods eliminate
distributions from the differential equations based on predictions for local
dynamical effects of the distributions. Such effects are simply included
into the corresponding substitutions so that new equations are free of
singular terms and thus satisfy the so-called Caratheodory conditions. As a
result, it is possible to prove existence and investigate different
properties of solutions. Such a generalization is based on the integral form
of the differential equation $\dot{x}=f(t,x)$ with a continuous right-hand
side

\begin{equation}
x(t)=x(t_{0})+\dint\limits_{0}^{t}f(s,x(s))ds  \label{integral form}
\end{equation}

If the function $f(t,x)$ is discontinuous in $t$, but still continuous in $x$%
, then the functions satisfying (\ref{integral form}) can be considered as
solutions of the equation $\dot{x}=f(t,x)$. Generally, integration in (\ref%
{integral form}) should comply with the concept of Lebesgue integral. In
this case, the function $f(t,x)$ does not have to be point-wise defined%
\footnote{%
Practically, such an extension almost never contradicts to the physical
contents of modeling; recall that the differential equations of motion are
derived from variational principles formulated in the integral form.}.
However, the following Caratheodory conditions must be satisfied: In the
domain $D$ of the $(t,x)$-space, the function $f(t,x)$ is 1) defined and
continuous in $x$ for almost all $t$, 2) measurable in $t$ for each $x$, and
3) the estimate $|f(t,x)|\leq m(t)$ holds, where $m(t)$ is a summable
function on each finite interval, if $t$ is not bounded in $D$.

As a simplified illustration, let us consider a single degree-of-freedom
system whose velocity, $v=v(t)$, is described by the differential equation%
\begin{equation}
\dot{v}+kv^{3}=q\delta \left( t-t_{1}\right)  \label{AP1}
\end{equation}%
where $k$, $q$ and $t_{1}$ are constant parameters, and $\delta $ is the
Dirac'delta function.

Equation (\ref{AP1}) describes a unit-mass particle in a nonlinearly viscous
media with the cubic dissipation law. In this case, the $\delta $-input
generates a step-wise discontinuity of the response $v(t)$ at $t=t_{1}$,
nevertheless the nonlinear operation in (\ref{AP1}) remains meaningful.
Moreover, the $\delta $-pulse can be eliminated from equation (\ref{AP1}) as
follows. Since the velocity must be bounded, then only the inertia term can
compensate the impact force on the right-hand side of equation (\ref{AP1}).
In other words, near the impact time $t=t_{1}$, equation (\ref{AP1}) is
linearized as $\dot{v}=q\delta \left( t-t_{1}\right) $. Comparing this with
equation (\ref{theta dot}), gives solution 
\begin{equation}
v=u+q\theta \left( t-t_{1}\right)  \label{AP2}
\end{equation}%
where $u$ is an arbitrary constant.

Let us assume now that $u$ is a function of time $u=u\left( t\right) $ such
that solution (\ref{AP2}) becomes valid for the original equation (\ref{AP1}%
) on the entire time interval, $-\infty <t<\infty $. Then, substituting (\ref%
{AP2}) in (\ref{AP1}), and taking into account that $\dot{\theta}%
(t-t_{1})=\delta \left( t-t_{1}\right) $ and $\theta ^{2}\left(
t-t_{1}\right) =\theta \left( t-t_{1}\right) $, gives%
\begin{equation}
\dot{u}+k[u^{3}+(q^{3}+3uq^{2}+3u^{2}q)\theta \left( t-t_{1}\right) ]=0
\label{u dot}
\end{equation}

In contrast to (\ref{AP1}), equation (\ref{AP2}) includes no $\delta $%
-function and thus admits visualization of its phase flow on the $(t,v)$%
-plane with the related qualitative study.

Note that substitution (\ref{AP2}) generalizes on equation%
\begin{equation}
\dot{v}=f(v,t)+\dsum\limits_{i=1}^{\infty }q_{i}\delta \left( t-t_{i}\right)
\label{AP1 general}
\end{equation}%
where $f(v,t)$ is assumed to have no singularities within some domain of the 
$(v,t)$-plane, and $\{q_{i}\}$ and $\{t_{i}\}$ are constants.

In this case, the following substitution eliminates all the $\delta $%
-functions from equation (\ref{AP1 general})%
\begin{equation}
v(t)=u(t)+\dsum\limits_{i=1}^{\infty }q_{i}\theta \left( t-t_{i}\right)
\label{AP2 general}
\end{equation}

Further generalization on the vector-form equations is quite obvious.
Complications may occur however when the structure of original equations is
changed as described in the next section.

\subsection{Distributions as parametric inputs in dynamical systems}

This specific case illustrates the major issue of using distributions for
modeling dynamical systems. Following reference \cite{Filippov:1988Kluwer},
consider the linear initial value problem%
\begin{eqnarray}
\dot{v}+k\delta \left( t-t_{1}\right) v &=&0  \label{param1} \\
v(0) &=&v_{0}  \notag
\end{eqnarray}%
where $k$ and $t_{1}>0$ are constant.

Since $\dot{v}=0$ for $t<t_{1}$ and $t>t_{1}$, then%
\begin{equation}
v=v_{0}[1-\lambda \theta \left( t-t_{1}\right) ]  \label{param2}
\end{equation}%
where $\lambda $ is yet unknown constant parameter of discontinuity such
that $v=v_{0}$ for $t<t_{1}$ and $v=-\lambda v_{0}\ $for $t>t_{1}$.

Equation (\ref{param1}) shows that the jump of the solution depends on the
behavior of solution itself near the point $t=t_{1}$. In this case, there is
no certain choice for $\lambda $ since its magnitude depends upon additional
assumptions regarding the model \cite{Filippov:1988Kluwer}. To clarify the
details, let us substitute (\ref{param2}) in (\ref{param1}) with the intent
to find $\lambda $. This gives%
\begin{equation}
(k-\lambda )\delta (t-t_{1})-k\lambda \theta (t-t_{1})\delta (t-t_{1})=0
\label{param3}
\end{equation}

It was mentioned in Introduction that the combination $\theta
(t-t_{1})\delta (t-t_{1})$ has no certain meaning in the distribution
theory. However, the form of (\ref{param2}) together with equation (\ref%
{param1}) dictate%
\begin{equation}
\theta (t-t_{1})\delta (t-t_{1})=\alpha \delta (t-t_{1})  \label{param4}
\end{equation}%
and therefore,%
\begin{equation}
\lambda =\frac{k}{1+\alpha k}  \label{param6}
\end{equation}%
where the magnitude of parameter $\alpha $ depends on the model' assumptions.

For instance, the number $\alpha =1/2$ was obtained in introductory example
(1). In the present case, it is unknown whether the functions $\theta
(t-t_{1})$ and $\delta (t-t_{1})$ are generated by the same family of smooth
functions. Nevertheless, physical approaches to determining $\alpha $ may be
similar to that described in Introduction. Namely, let us replace the Dirac
function $\delta (t-t_{1})$ in equation (\ref{param1}) by its smooth
preimage $\delta _{\varepsilon }(t-t_{1})$ defined in Introduction. As a
result, equation (\ref{param1}) takes the form of a regular separable
equation whose solution is%
\begin{equation}
v=v_{0}\exp \left[ -\frac{1}{2}k\left( \tanh \frac{1}{\varepsilon ^{2}}%
+\tanh \frac{t-t_{1}}{\varepsilon ^{2}}\right) \right]  \label{param5}
\end{equation}

This gives%
\begin{equation}
v=v_{0}\left\{ 1-[1-\exp (-k)]\theta (t-t_{1})\right\} \text{\quad as\quad }%
\varepsilon \rightarrow 0  \label{param7}
\end{equation}

Comparing (\ref{param7}) to (\ref{param2}) and taking into account (\ref%
{param6}), gives $\lambda =1-\exp (-k)$ and therefore%
\begin{equation}
\alpha =\frac{1}{1-\exp (-k)}-\frac{1}{k}  \label{param8}
\end{equation}

Expression (\ref{param8}) shows that definition (\ref{param4}) depends upon
the model parameter $k$; see Fig. 2 for illustration. In particular, $\alpha
\rightarrow 1/2$ as $k\rightarrow 0$, and this brings us back to the case
considered in Introduction.

Finally, note that substitutions of type (\ref{AP2}) and (\ref{AP2 general})
as well as its different variations are widely used in the literature to
describe moving discontinuity waves \cite{Whitham:1999Wiley}, \cite%
{Maslov:1981UMN}, \cite{Haller:2008}. In most such cases though there is no
explicit source of discontinuities that naturally occur as a result of wave
shape evolutions predetermined by initial perturbations and inner properties
of wave model.

\subsection{Nonsmooth positional coordinates}

The idea of nonsmooth coordinates associates with elastic but perfectly
stiff barriers reflecting moving particles in a mirror-wise manner. Since
outcome of such reflections is predictable then it can be built into the
mechanical model in advance through the corresponding nonsmooth coordinate
transformations. It was shown in references \cite{Zhuravlev:1976MTT}, \cite%
{Zhuravlev:1978PMM}, and \cite{Zhuravlev:1988Nauka} that introducing
nonsmooth coordinates simply eliminates barriers by unfolding the
configuration space to include the area behind the barrier. As a result, the
differential equations of motion are derived on the entire time interval
with no need in formulation of impact conditions. For illustrating purposes,
consider the following $N$-degree-of-freedom Lagrangian system%
\begin{eqnarray}
L &=&\frac{1}{2}\dsum\limits_{i=1}^{N}\dot{q}_{i}^{2}-\dsum%
\limits_{i=0}^{N}k_{i}(q_{i+1}-q_{i})^{2}  \label{Lagrangian} \\
|q_{i}(t)| &\leq &1  \label{constraint} \\
q_{0}(t) &\equiv &q_{N+1}(t)\equiv 0  \label{bound cond}
\end{eqnarray}

This is a chain of unit-mass particles connected by linearly elastic springs
of stiffness $k_{i}$. Perfectly stiff elastic constraints are imposed on
each of the coordinates according to (\ref{constraint}). Although Lagrangian
(\ref{Lagrangian}) generates linear differential equations, these equations
alone do not describe the entire system. Due to the presence of constraints (%
\ref{constraint}), the system is actually strongly nonlinear that becomes
obvious in adequately chosen coordinates. Transition to such coordinates is
described by%
\begin{equation}
q_{i}=\tau (x_{i})  \label{zhurav subst}
\end{equation}%
where $\tau $ is the triangular sine-wave

\begin{eqnarray}
\tau \left( x\right) &=&\frac{2}{\pi }\mathrm{\arcsin \sin }\frac{\pi x}{2}%
=\left\{ 
\begin{array}{ll}
x & \text{for }-1\leq x\leq 1 \\ 
-x+2 & \text{for }1\leq x\leq 3%
\end{array}%
\right.  \label{tau} \\
&&\tau \left( x\right) \overset{\forall x}{=}\tau \left( 4+x\right)  \notag
\end{eqnarray}

Note that notation (\ref{tau}) and normalization of the period differ from
those introduced in original works \cite{Zhuravlev:1976MTT}, \cite%
{Zhuravlev:1978PMM}, \cite{Zhuravlev:1988Nauka}. The only reason for such
modification is to deal with the triangular wave of unit slope\footnote{%
Although this condition does no matter for the method introduced in
references \cite{Zhuravlev:1976MTT}, \cite{Zhuravlev:1978PMM}, \cite%
{Zhuravlev:1988Nauka}, Section 3 of the present paper describes another
method for which normalization (\ref{tau der}) is essential.}%
\begin{equation}
\lbrack \tau ^{\prime }(x)]^{2}=1  \label{tau der}
\end{equation}%
for at least almost all $x$.

The coordinate transformation (\ref{zhurav subst}) brings system (\ref%
{Lagrangian}) through (\ref{bound cond}) to the form%
\begin{eqnarray}
L &=&\frac{1}{2}\dsum\limits_{i=1}^{N}\dot{x}_{i}^{2}-\dsum%
\limits_{i=0}^{N}k_{i}[\tau (x_{i+1})-\tau (x_{i})]^{2}  \label{Lagrangian s}
\\
x_{0}(t) &\equiv &x_{N+1}(t)\equiv 0  \label{bound cond s}
\end{eqnarray}

It is seen from (\ref{Lagrangian s}) that transformation (\ref{tau})
preserves the quadratic form of kinetic energy while the constraint
conditions (\ref{constraint}) are satisfied automatically due to the
property $|\tau \left( x\right) |\leq 1$. In contrast to (\ref{Lagrangian}),
Lagrangian (\ref{Lagrangian s}) completely describes the model on the entire
time interval $0\leq t<\infty $. However, in terms of the new coordinates $%
x_{i}$, the potential energy acquired a non-local cell-wise structure so
that the corresponding differential equations of motion are essentially
nonlinear; for example, see (\ref{eqns transformed 2}) below. Now every
impact interaction with constraints is interpreted as a transition from one
cell to another as illustrated below on the two degrees-of-freedom model, $%
N=2$. In this case, Lagrangian (\ref{Lagrangian s}) gives the differential
equations of motion on the infinite plane $-\infty <x_{i}<\infty $ ($i=1,2$)
with no constraints 
\begin{eqnarray}
\ddot{x}_{1}+[(k_{0}+k_{1})\tau (x_{1})-k_{1}\tau (x_{2})]\tau ^{\prime
}(x_{1}) &=&0  \notag \\
\ddot{x}_{2}+[(k_{1}+k_{2})\tau (x_{2})-k_{1}\tau (x_{1})]\tau ^{\prime
}(x_{2}) &=&0  \label{eqns transformed 2}
\end{eqnarray}

Fig. 3 shows the corresponding equipotential energy levels and a sample
trajectory of beat-wise dynamics represented by Figs.4 and 5 in the original
coordinates. Fig. 3 shows, for instance, that the system is trapped in some
cells for the energy exchange process. After one of the two masses
accumulated the energy, which is sufficient to reach the barrier, the impact
event happens accompanied transition to another cell. The fact of energy
exchange inside a trapping cell is confirmed by the transversality of
incoming and outcoming pieces of the trajectory. As long as the mass remains
in impact regime, its trajectory is passing through one cell to another
until the system is trapped again in rather another cell for a new energy
exchange process. A similar geometrical interpretation but for impact mode
dynamics was introduced earlier in \cite{Vedenova:1985PMM}, where the impact
modes were associated with `hidden geometrical symmetries' revealed by
periodic patterns of equipotential lines as shown in Fig. 6.

In particular, closed form analytical solutions for different impact modes
were obtained by means of the averaging procedure. Note that, according to
the original works \cite{Zhuravlev:1976MTT}, \cite{Zhuravlev:1978PMM}, \cite%
{Zhuravlev:1988Nauka}, applicability of the averaging procedure constitutes
the major advantage given by transformation (\ref{zhurav subst}) since
infinite impact forces are effectively eliminated from the system.

Similar kind of visualization for a two-degree-of-freedom vibrating system
with only one mass under two-sided constraint condition was used in \cite%
{Pilipchuk:2000ND}.

Let us consider a one-degree-of-freedom unit mass oscillator with the
potential energy of restoring force $P\left( q\right) $, 
\begin{equation}
L=\frac{1}{2}\dot{q}^{2}-P\left( q\right)  \label{A1}
\end{equation}%
whose motion is limited by the interval 
\begin{equation}
-1\leq q  \label{A1_cond}
\end{equation}

It is assumed that the oscillator collides with a one-sided perfectly stiff
barrier at $q=-1$ with no energy loss. In this case, the constraint (\ref%
{A1_cond}) is eliminated by the space unfolding coordinate transformation $%
q\rightarrow x$: 
\begin{equation}
q=-1+|x+1|  \label{oneside q2x}
\end{equation}%
where the constant shifts in (\ref{oneside q2x}) are chosen to preserve the
origin and barrier positions.

Substituting (\ref{oneside q2x}) in (\ref{A1}), gives 
\begin{equation}
L=\frac{1}{2}\dot{x}^{2}-P(|x+1|-1)  \label{A5}
\end{equation}

The system' phase trajectories are described by the energy integral $\dot{x}%
^{2}/2+P(|x+1|-1)=Const.$ This family of curves is illustrated by Fig. 7 on
the phase plane $x\dot{x}$ at different energy levels in the case $%
P(q)=q^{2}/2$. In particular, it is seen that high-energy trajectories are
non-smooth due to mirror-wise reflections against the barrier at $x=-1$ ($%
q=-1$). As a result, the phase portrait of the oscillator has effectively
non-local structure, where any transition from one side of the plane to
another is the result of impact interaction with the barrier.

Although original works \cite{Zhuravlev:1976MTT}, \cite{Zhuravlev:1978PMM}, 
\cite{Zhuravlev:1988Nauka}, deal with illustrating models of deterministic
dynamics, further applications were shifted mostly into the area of random
vibrations may be due to earlier works \cite{Dimentberg:1988Wiley} and \cite%
{Dimentberg:1988RSP}. Different analytical and numerical tools in this area
were developed during recent few years; see references \cite%
{Namachchivaya:2005JAM}, \cite{Dimentberg:2005IJBC}, \cite%
{Dimentberg:2009IJNM}, \cite{Rong:2009JSV}. From the standpoint of practical
applications, inelastic effects of interactions with stiff constraints
become essential. Generally, impact dissipation effects can be modeled by
the dissipative term \cite{Zhuravlev:1988Nauka}, \cite{Babitsky:1998Springer}%
, \cite{Dimentberg:2009IJNM} $(1-\kappa )\dot{x}|\dot{x}|\delta _{-}(x)$,
where $\delta _{-}(x)$ is a specific rule rather than conventional Dirac
function. According to this rule, the impulsive damping acts right before
the result of such damping namely velocity jump occurs. Such damping model
is justified if the restitution coefficient $\kappa $ is close to unity so
that the factor $1-\kappa $ is small. In this case, the integral effect of
the impulsive damping can play the role perturbation within asymptotic
procedures, in which the velocity $\dot{x}$ is given by an unperturbed
system and therefore remains continuous. Distributed viscous damping effects
still can be described in a regular way by continuos terms as it was done,
for instance, in \cite{Avramov:2008IAM}. Non-elastic impact interactions
with constraints can be modeled also in a purely geometrical way under some
conditions on the class of motions though \cite{Zhuravlev:1988Nauka}.

Another approach that deals with the class of nonsmooth `non-conservative'
transformations is described in the next subsection.

\subsection{Nonsmooth transformation of state variables}

Generalized approaches to eliminating non-elastic constraints should
obviously involve both types of the state variables - coordinates and
velocities. For illustrating purposes, let us consider the case of harmonic
oscillator under the constraint condition 
\begin{equation}
\mathbf{\dot{x}}=\mathbf{Ax}  \label{iv1}
\end{equation}%
\begin{equation}
x_{1}>0  \label{ivConstr}
\end{equation}%
where $\mathbf{x}=[x_{1}(t),x_{2}(t)]^{T}$ is the system' state vector such
that $x_{2}=\dot{x}_{1}$, and 
\begin{equation}
\mathbf{A}=\left[ 
\begin{array}{ll}
0 & 1 \\ 
-\omega ^{2} & 0%
\end{array}%
\right]  \label{iv2}
\end{equation}%
\qquad

It is also assumed that every collision with the constraint at $x_{1}=0$ is
accompanied by a momentary energy loss characterized by the coefficient of
restitution $\kappa $, in other words

\begin{equation}
x_{2}(t^{\ast }+0)=-\kappa x_{2}(t^{\ast }-0)  \label{ivImpact}
\end{equation}%
where $t^{\ast }$ is the collision time, at which $x_{1}($ $t^{\ast })=0$.

The idea is to unfold the phase space in such way that the energy loss
occurs automatically whenever the system crosses preimage of the line $%
x_{1}=0$. The corresponding non-conservative transformation was introduced
in \cite{Ivanov:1993PMM}, \cite{Ivanov:1994JSV}, \cite%
{Ivanov:1997International} as a transformation of state vector, $\mathbf{%
x\longrightarrow y}$, of the form 
\begin{equation}
\mathbf{x}=\mathbf{Sy}  \label{iv3}
\end{equation}%
where $\mathbf{y}=[s(t),v(t)]^{T}$ is a new state vector, and the transition
matrix is given by%
\begin{equation}
\mathbf{S}=\left[ 
\begin{array}{ll}
1 & 0 \\ 
0 & 1-k\text{sgn}(sv)%
\end{array}%
\right] \text{sgn}(s)  \label{iv4}
\end{equation}%
where $k=(1-\kappa )/(1+\kappa )$.

Note that transformation (\ref{iv3}) is strongly nonlinear due to the
nonsmooth dependence $\mathbf{S=S(y},k\mathbf{)}$. Nevertheless,
substitution (\ref{iv3}), gives equation 
\begin{equation}
\mathbf{\dot{y}=(S}^{-1}\mathbf{AS)y}  \label{iv5}
\end{equation}

There is some `hidden issue'\ with substitution (\ref{iv3}) since the result
(\ref{iv5}) has the same form as it would have in the case of constant
matrix $\mathbf{S}$. In fact, the matrix $\mathbf{S}$ is constant just
almost everywhere. A formal substitution of (\ref{iv3}) in (\ref{iv1}) would
eventually impose specific conditions on distributions similar to (\ref%
{sgn*delta}).

In the component-wise form, expressions (\ref{iv3}) and (\ref{iv5}) are
written as, respectively, 
\begin{eqnarray}
x_{1} &=&x_{1}(s,v)\equiv s\text{sgn}(s)  \notag \\
x_{2} &=&x_{2}(s,v)\equiv \text{sgn}(s)[1-k\text{sgn}(sv)]v  \label{iv7}
\end{eqnarray}%
and 
\begin{eqnarray}
\dot{s} &=&[1-k\text{sgn}(sv)]v  \notag \\
\dot{v} &=&-\omega ^{2}s[1+k\text{sgn}(sv)]/(1-k^{2})  \label{iv6}
\end{eqnarray}

Now both unknown components of the state vector are continuous, whereas
effects of non-elastic collisions (\ref{ivImpact}) are captured by
transformation (\ref{iv7}).

Finally, consider the general case of one-degree-of-freedom nonlinear
oscillator 
\begin{eqnarray}
\dot{x}_{1} &=&x_{2}  \notag \\
\dot{x}_{2} &=&-f(x_{1},x_{2},t)  \label{iv8}
\end{eqnarray}%
whose motion is restricted to the positive half plane $x_{1}>0$ by a
non-elastic barrier at $x_{1}=0$ of the restitution coefficient $\kappa $.

Applying transformation (\ref{iv7}) to system (\ref{iv8}), gives 
\begin{eqnarray}
\dot{s} &=&[1-k\text{sgn}(sv)]v  \notag \\
\dot{v} &=&-f(x_1(s,v),x_2(s,v),t)\text{sgn}(s)[1+k\text{sgn}(sv)]/(1-k^2)
\label{iv9}
\end{eqnarray}

Although the technique is illustrated on a one-degree-of-freedom model,
similar coordinate transformations apply to multiple degree-of-freedom
systems by choosing one of the coordinates perpendicular to the constraint.
For that reason, it is convenient to use the Routh descriptive function
whose normal to the constraint coordinate is Lagrangian whereas other
coordinates and associated momenta are Hamiltonian \cite{Ivanov:1994JSV}.

Finally, note that adjustments of classes of smoothness of dynamical systems
by eliminating infinite discontinuities extend the set of applicable
analytical tools. For instance, problems of stability and bifurcation
analyses of impact motions were considered in \cite{Ivanov:1994JSV} by means
of the linearization technique developed for dynamical systems with
nonsmooth right-hand sides \cite{Aizerman:1958PMM}. In particular, the
fundamental matrix of explicit form and the corresponding characteristic
equations were obtained. As a result, investigation of stability and
bifurcations became possible by using conventional methods. In fact, before
the transformation, discontinuities of phase trajectories as those shown in
Fig. 9 would complicate any local analyses. However, the transformation
improves the class of smoothness to the extent which is needed to build
major objects of local analyses and averaging tools. A regular approach to
stability and bifurcation analysis in impact systems was proposed in \cite%
{Ivanov:1996CSF}. In particular, it was shown that the discontinuous
bifurcation of grazing impact can be regularized. This may lead to a new
interpretation of grazing bifurcations. Namely, after such bifurcation, some
periodic motion might survive and even preserve stability.

\section{Nonsmooth temporal arguments}

In this section, we describe nonsmooth substitutions of the independent
variable, which is temporal argument in the present text. It will be shown
below that such nonsmooth substitutions associate with some common temporal
symmetries of motions regardless of types of systems. This approach was
originally developed for the class of strongly non-linear but smooth
oscillators \cite{Pilipchuk:1985PMM}, \cite{Pilipchuk:1988DANUkrUkr}.
However, its main specifics are clearly seen even in the case of
non-oscillatory motion of a classic unit-mass particle under returning
potential force as shown in Figs. 10 and 11. \textit{The idea is to employ
most elementary macrodynamical processes whose specifics nevertheless
provide sufficient conditions for observing strong nonlinearities} \cite%
{Pilipchuk:2010Springer}. Examples of such `elementary strongly nonlinear
processes' are found among the rigid-body motions as those shown in Fig. 12.
However, the key question is how to bridge the gap between the classes of
smooth and nonsmooth motions within the same mathematical formalism. It will
be shown below on simple examples that nonsmooth substitutions of temporal
argument may play the role of such a bridge.

\subsection{Positive time}

Let the potential energy $P(x)$ be a smooth function of the time dependent
coordinate $x$ as qualitatively shown in Fig. 10. Then the differential
equation of motion of a unit-mass particle under the force $f(x)=P^{\prime
}(x)$ is given by%
\begin{equation}
\ddot{x}+f(x)=0  \label{nstt1}
\end{equation}

The initial conditions are $x=x_{0}>0$ and $\dot{x}=v_{0}<0$ at $t=t_{0}<0$
as shown in Figs. 10 and 11.

As the particle reaches a turning point at some time $t=a$ it makes a
U-turn. Since equation (\ref{nstt1}) admits the group $t\longrightarrow -t$,
the reverse motion will be symmetric with respect to the U-turn point $t=a$.
Such a `prediction' builds into the differential equation of motion (\ref%
{nstt1}) through the new temporal argument $t\longrightarrow s$:%
\begin{equation}
s=|t-a|\text{,\quad }x=x(s)  \label{nstt2}
\end{equation}

The mechanical model generating such time argument by its natural motion is
shown in Fig. 12 (a). Since at almost all $t$,%
\begin{equation}
\dot{s}^{2}=1  \label{nstt3}
\end{equation}%
then, substituting (\ref{nstt2}) in (\ref{nstt1}), gives%
\begin{equation}
\frac{d^{2}x}{ds^{2}}+f(x)=0\text{,\quad }s>0  \label{nstt4}
\end{equation}%
under condition%
\begin{equation}
\frac{dx}{ds}=0\text{\quad if\quad }s=0  \label{nstt5}
\end{equation}

It is easy to see that the boundary condition (\ref{nstt5}) eliminates
singularity of second time derivative that formally occurs due to the
non-smoothness of substitution (\ref{nstt2}).

As follows from Fig. 11, substitution (\ref{nstt2}) reverses the time
direction exactly when the particle makes a U-turn. Although the form of the
equation remains the same, substitution (\ref{nstt2}) gives certain
advantages from both physical and mathematical standpoints. First, equation %
\ref{nstt4}, even after being drastically simplified as $d^{2}x/ds^{2}=0$,
still preserves the major dynamical event, which is the U-turn of the
particle. In this degenerated case, the smooth potential barrier is
effectively replaced by a perfectly stiff one as follows from the general
solution, $x=As(t)+B$, where $A$ and $B$ are arbitrary constants of
integration. Now, if some perturbation series converges for $s\geq 0$, then
it is automatically converges for the entire interval of the original time, $%
-\infty <t<\infty $.

As another example, let us consider the case of impulsively loaded single
degree-of-freedom system,%
\begin{equation}
\ddot{x}+f(x)=p\ddot{s}  \label{nstt6}
\end{equation}%
where $\ddot{s}=2\delta (t-a)$ and $p=const$.

Substituting (\ref{nstt2}) in (\ref{nstt6}) and taking into account (\ref%
{nstt3}), gives

\begin{equation}
\frac{d^{2}x}{ds^{2}}+f(x)=\left( p-\frac{dx}{ds}\right) \ddot{s}
\label{nstt7}
\end{equation}

Eliminating the singularity $p\ddot{s}$ in (\ref{nstt7}), gives the same
equation (\ref{nstt4}) however under non-homogeneous boundary condition%
\begin{equation}
\frac{dx}{ds}=p\text{\quad if\quad }s=0  \label{nstt8}
\end{equation}

Since substitution (\ref{nstt2}) is non-invertible in the form $t=t(s)$ on
the entire time interval, then using the argument $s$ in general case of
dynamical system appears to be less straightforward but nevertheless
possible based on the following identity%
\begin{equation}
t=a+s\dot{s}  \label{nstt9}
\end{equation}

Due to relationship (\ref{nstt3}), the combination (\ref{nstt9}) represents
a specific complex number with the basis $\{1,\dot{s}\}$ \cite%
{Pilipchuk:2005IJBC}. In contrast to the conventional elliptic complex
algebra, the operation $1/t$ with (\ref{nstt9}) may not hold. Interestingly
enough, such algebraic structures has been known for quite a long time \cite%
{Cockle:1848}, \cite{Clifford:1882} mostly as abstract mathematical objects
with no relation to nonsmooth functions or impact systems. In the modern
mathematical literature, these numbers are often referred to as a simple
example of Clifford algebras under the name `hyperbolic algebra' \cite%
{Sobczyk:1995CMJ} with very many synonyms though \cite{Wikipedia:Hyperbolic}%
. Some areas of physics are linked to this algebra quite closely \cite%
{Hucks:1993JMP}, but constructive applications are rather limited.

Let us formulate useful algebraic properties of hyperbolic numbers adapted
to the form (\ref{nstt9}). For instance, the hyperbolic conjugate to (\ref%
{nstt9}) is introduced as $t^{-}=a-s\dot{s}$, and the hyperbolic modulus is
calculated as $|t|_{h}=\sqrt{|tt^{-}|}=\sqrt{|a^{2}-s^{2}|}$. The term 
\textit{hyperbolic} associates with the fact that the relationship $%
|t|_{h}=\rho $ with some fixed $\rho $ describes a four branched hyperbola
on the hyperbolic plane with the basis $\{1,\dot{s}\}$:%
\begin{eqnarray*}
t &=&\pm \rho (\cosh \phi +\dot{s}\sinh \phi )\equiv \pm \rho \exp (\dot{s}%
\phi )\text{,}\quad \phi =\text{arctanh}(s/a) \\
t &=&\pm \rho (\sinh \phi +\dot{s}\cosh \phi )\equiv \pm \rho \exp (\dot{s}%
\phi )\text{,}\quad \phi =\text{arctanh}(a/s)
\end{eqnarray*}

The hyperbolic numbers create isomorphism with symmetric $2\times 2$%
-matrixes, so that, for instance%
\begin{equation}
(a+s\dot{s})^{2}\longleftrightarrow \left( 
\begin{array}{cc}
a & s \\ 
s & a%
\end{array}%
\right) ^{2}  \label{nstt10}
\end{equation}

Note that, in our case, the hyperbolic structure is generated by time
dependent transformations rather than imposed on abstract elements by
mathematical definitions. As a result, differential and integral properties
of hyperbolic numbers can be introduced \cite{Pilipchuk:2005IJBC}. So, for
practically any function $x\left( t\right) $, it can be shown that%
\begin{equation}
x\left( t\right) =x\left( a+s\dot{s}\right) =X\left( s\right) +Y\left(
s\right) \dot{s}  \label{nstt11}
\end{equation}%
where%
\begin{eqnarray}
X\left( s\right) &=&\frac{1}{2}\left[ x\left( a+s\right) +x\left( a-s\right) %
\right]  \label{nstt12} \\
Y\left( s\right) &=&\frac{1}{2}\left[ x\left( a+s\right) -x\left( a-s\right) %
\right]  \notag
\end{eqnarray}

Then, taking into account (\ref{nstt3}), gives%
\begin{equation}
\dot{x}\left( t\right) =Y^{\prime }\left( s\right) +X^{\prime }\left(
s\right) \dot{s}+p\ddot{s}  \label{nstt13}
\end{equation}%
where%
\begin{equation}
Y\left( 0\right) =p  \label{nstt14}
\end{equation}

If $x(t)$ is continuous then obviously $p=0$ and derivative (\ref{nstt13})
also belongs to the hyperbolic algebra, otherwise the quantity $p$ may serve
for elimination of singularities as seen from equation (\ref{nstt7}).

As a simple application, consider the initial value problem%
\begin{eqnarray}
\dot{x}+\lambda x &=&2p\delta (t-a)=p\ddot{s}  \label{nstt15} \\
x(0) &=&0  \notag
\end{eqnarray}%
where $\lambda $ and $p$ are constant.

By considering functions $X\left( s\right) $ and $Y\left( s\right) $ in (\ref%
{nstt11}) as new unknowns, then substituting (\ref{nstt11}) through (\ref%
{nstt13}) in (\ref{nstt15}), and taking into account the linear independence
of basis $\{1,\dot{s}\}$, gives%
\begin{eqnarray}
Y^{\prime }+\lambda X &=&0  \label{nstt16} \\
X^{\prime }+\lambda Y &=&0  \notag \\
X(a)-Y(a) &=&0  \notag
\end{eqnarray}%
under condition (\ref{nstt14}).

The boundary value problem (\ref{nstt14}) is free of $\delta $-functions and
admits an obvious solution , $X\equiv Y=p\exp (-\lambda s)$. Substituting
these $X$- and $Y$-components in (\ref{nstt11}), gives solution of the
initial value problem (\ref{nstt15}) in the form%
\begin{equation}
x=p\exp (-\lambda s)(1+\dot{s})  \label{nstt17}
\end{equation}

The link established between nonsmooth temporal substitution (\ref{nstt9})
and the hyperbolic structure essentially facilitates analytical
manipulations with differential equations. For example, instead of using the
standard basis $\{1,\dot{s}\}$ it is possible to introduce the so-called
idempotent basis, associated with the two isotropic lines which separate the
hyperbolic quadrants as follows 
\begin{eqnarray}
i_{\pm } &=&\frac{1}{2}(1\pm \dot{s})  \label{nstt18} \\
i_{\pm }^{2} &=&i_{\pm }\text{,\quad }i_{+}i_{-}=0  \notag
\end{eqnarray}

Transition to the idempotent basis in (\ref{nstt9}) gives%
\begin{equation}
t=(a+s)i_{+}+(a-s)i_{-}  \label{nstt19}
\end{equation}

Due to the properties of basis (\ref{nstt18}), substitution (\ref{nstt19})
possesses `functional linearity.' In other words,%
\begin{equation}
t^{\alpha }=(a+s)^{\alpha }i_{+}+(a-s)^{\alpha }i_{-}  \label{nstt20a}
\end{equation}%
for any real $\alpha $, and generally, 
\begin{eqnarray}
x(t) &=&x[(a+s)i_{+}+(a-s)i_{-}]  \notag \\
&=&x(a+s)i_{+}+x(a-s)i_{-}  \label{nstt20} \\
&\equiv &X_{+}(s)i_{+}+X_{-}(s)i_{-}  \notag
\end{eqnarray}

The advantage of the idempotent basis is that the differential equations of
motion with respect to components $X_{+}$ and $X_{-}$ are decoupled. Instead
the boundary conditions become coupled though. Different examples of
solutions in the idempotent basis are given in recent publications \cite%
{Pilipchuk:2009ND} and \cite{Pilipchuk:2010Springer}.

\subsection{Triangular sine-wave time substitution}

Since any vibrating process is a sequence of U-turns then the corresponding
nonsmooth time substitution can be combined of functions given by (\ref%
{nstt2}) with different signs and temporal shifts. In periodic case of the
period $T=4$, such combination is given by the sawtooth function (\ref{tau}%
), whose argument is replaced by time, $\tau =\tau (t)$. A mechanical model
generating such time substitution by its free motion is shown in Fig. 12 (b)
while the periodic version of identity (\ref{nstt9}) is 
\begin{equation}
t=1+\left( \tau -1\right) \dot{\tau}\qquad \text{if }-1<t<3  \label{nstt21}
\end{equation}%
where $\dot{\tau}^{2}=1$, and therefore (\ref{nstt21}) is a hyperbolic
number with the basis $\{1,\dot{\tau}\}$.

In physical terms, it follows from (\ref{nstt21}) that any periodic process,
whose period is normalized to $T=4$, is uniquely expressed through the
dynamic states of standard impact oscillator in the form \cite%
{Pilipchuk:2005IJBC} 
\begin{equation}
x\left( t\right) =X\left( \tau \right) +Y\left( \tau \right) \dot{\tau}
\label{nstt22}
\end{equation}%
where 
\begin{eqnarray}
X\left( \tau \right) &=&\frac{1}{2}\left[ x\left( \tau \right) +x\left(
2-\tau \right) \right]  \notag \\
Y\left( \tau \right) &=&\frac{1}{2}\left[ x\left( \tau \right) -x\left(
2-\tau \right) \right]  \label{nstt23}
\end{eqnarray}

Identity (\ref{nstt22}) means that the triangular sine and rectangular
cosine waves capture temporal symmetries of periodic processes regardless
specifics of individual vibrating systems.

Introducing a slow temporal scale in (\ref{nstt22}) extends the area of
applications on modulated vibrating processes. In such cases, two variable
expansions \cite{Kevorkian:1996Springer} can be used by considering the
triangular sine time as a fast scale \cite{Pilipchuk:2010Springer}.

Different applications of nonsmooth argument substitutions with the related
techniques to problems of theoretical and applied mechanics can be found in 
\cite{Pilipchuk:1997PMM}, \cite{Manevitch:2007ArchAM}, \cite%
{Manevitch:2006Samos}, \cite{Manevitch:2008PIME}, \cite{Gendelman:2001ASME}, 
\cite{Vakakis:1999ASME}, \cite{Lee:2005PhysD}, \cite{Lee:2009PhysD}, \cite%
{Sheng:2006MMT}, \cite{Vakakis:2002IJNM}, \cite{Salenger:1998IJNM}, \cite%
{Salenger:1998MRC}, \cite{Salenger:1999ND}, \cite{Starushenko:2002IJHMT}, 
\cite{Mikhlin:2000IJSS}, \cite{Mikhlin:2006MPE}, \cite{Thomsen:2008JSV}, 
\cite{Ibrahim:2009Springer}. The methodology was adapted also to the
nonlinear normal mode analyses and included in monograph \cite%
{Vakakis:1996Wiley}. While the idea of NNMs is effective in case of weak or
no energy exchange, the concept of the limiting phase trajectories \cite%
{Manevitch:2007ArchAM} considers the opposite situation namely intense
energy exchanges between weakly coupled oscillators or modes \cite%
{Manevitch:2006Samos}, \cite{Manevitch:2009arXiv}, \cite{Manevitch:2010PhysD}%
. In this case, nonsmooth time substitutions are invoked by the temporal
behavior of phase angle, which is responsible for energy distribution. This
resembles the sawtooth wave as the energy swing reaches its asymptotic limit.

Very recently, a class of strongly nonlinear traveling waves and localized
modes in one-dimensional homogeneous granular chains with no precompression
were considered in \cite{Starosvetsky:2011PhysRevE}. In particular, the
asymmetric version of identity (\ref{nstt22}) \cite{Pilipchuk:1999ND} was
applied. As a result, the authors developed a systematic semi analytical
approaches for computing different families of nonlinear traveling waves
parametrized by spatial periodicity (wave number) and energy.

\subsection{Quasi linear asymptotics for nonsmooth perturbations}

There are many physical and mechanical models described by linear
oscillators with small but nonsmooth perturbations. The related examples of
mechanical oscillators were considered in reference \cite{Nayfeh:1979Wiley}.
Nonsmooth but still continuous characteristics of elastic forces often occur
when modeling the dynamics of elastic structures with cracks \cite%
{Chen:1996ND}, \cite{Butcher:2007ND}, \cite{Vestroni:2007ND}. More
references and mathematical remarks on different cases of nonsmooth
perturbations can be found in reference \cite{Buika:2008arXiv}. As a
non-conservative case of nonsmooth perturbations, let us mention the
following modification of Van der Pol's oscillator \cite{Jordan:1999Oxford}, 
\cite{Hogan:2003JSV}

\begin{equation}
\ddot{q}+\varepsilon (|q|-1)\dot{q}+q=0  \label{vdp1}
\end{equation}%
where $\varepsilon $ is a small parameter.

Note that, using the nonsmooth term in (\ref{vdp1}), reduces the degree of
nonlinearity as compared to the classical Van der Pol's model. A periodic
limit cycle solution of equation (\ref{vdp1}) was obtained in \cite%
{Hogan:2003JSV} by taking into account the identity $|q|=$sgn$(q)q$ in order
to facilitate trigonometric expansions for the method of multiple scales.
The following generalized model was considered in \cite{Buika:2008arXiv} 
\begin{equation}
\ddot{q}+\varepsilon (|q|-1)\dot{q}+(1+\varepsilon \sigma )q=\varepsilon
\lambda \sin t  \label{vdp2}
\end{equation}%
where $\sigma $ is a detuning parameter.

Recall that classical methods of asymptotic integration require
perturbations to be as smooth as needed for deriving solutions of certain
asymptotic order. In such cases, non-smooth argument substitutions may
facilitate formulations for high-order asymptotic approximations. Moreover,
firts-order asymptotic solution are obtained exactly in a closed form. Let
us illustrate this remark on the one-degree-of-freedom oscillator 
\begin{equation}
\ddot{q}+\omega ^{2}q=\varepsilon \omega ^{2}\theta (q)q  \label{pwl1a}
\end{equation}%
where $\theta (q)$ is the Heaviside unit-step function.

The perturbation on the right-hand side of oscillator (\ref{pwl1a}) is a
continuous but non-smooth function of the coordinate $q$. Nevertheless, let
us represent the first-order asymptotic solution in the form%
\begin{eqnarray}
q &=&A\cos \varphi +\varepsilon q_{1}(\varphi )+O(\varepsilon ^{2})  \notag
\\
\varphi &=&\omega (1+\varepsilon \gamma _{1}+O(\varepsilon ^{2}))t
\label{asymp_sol}
\end{eqnarray}%
where $A=const.>0$, and $q_{1}$ and $\gamma _{1}$ are yet unknown
corrections to the generating solution obtained under the condition $%
\varepsilon =0$.

Substituting (\ref{asymp_sol}) in (\ref{pwl1a}) and matching first-order
terms of $\varepsilon $ on both sides of the equation, gives%
\begin{equation}
\frac{d^{2}q_{1}}{d\varphi ^{2}}+q_{1}=A\cos \varphi \lbrack 2\gamma
_{1}+\theta (\cos \varphi )]  \label{first_order_problem}
\end{equation}%
where the amplitude $A$ in the argument of unit-step function $\theta $ has
been ignored as a positive factor of no influence on the output.

For further comparison reason, let us reproduce first solution of equation (%
\ref{first_order_problem}) in terms of trigonometric expansions. According
to the idea of Poincare-Lindstedt method, the parameter $\gamma _{1}$ is
determined by using Fourier series with respect to $\varphi $ and
eliminating the resonance term on the right-hand side of equation (\ref%
{first_order_problem}) that gives $\gamma _{1}=-1/4$. The first-order
approximation $q_{1}(\varphi )$ is obtained then in the form%
\begin{eqnarray}
q &=&A\left[ \cos \varphi +\frac{\varepsilon }{\pi }\left( 1-\frac{2}{9}\cos
2\varphi +\frac{2}{225}\cos 4\varphi +...\right) \right] +O(\varepsilon ^{2})
\label{PL sol} \\
\varphi &=&\omega \left[ 1-\frac{\varepsilon }{4}+O(\varepsilon ^{2})\right]
t  \notag
\end{eqnarray}

The error of solution (\ref{PL sol}) depends upon both the magnitude of $%
\varepsilon $ and the number of terms retained in the Fourier series.

Now let us solve equation (\ref{first_order_problem}) by applying identity (%
\ref{nstt22}) to the right-hand side of equation (\ref{first_order_problem})
and its unknown $2\pi $-periodic solution. In other words, let us represent
solution in the form%
\begin{eqnarray}
q_{1} &=&X(\tau )+Y(\tau )e  \label{q1(phi)} \\
\tau &=&\tau (2\varphi /\pi )  \notag \\
e &=&e(2\varphi /\pi )\equiv d\tau (2\varphi /\pi )/d(2\varphi /\pi )  \notag
\end{eqnarray}%
where $X$ and $Y$ are new unknown functions of the argument $\tau $, and
obviously $e^{2}=1$ for almost all $\varphi $.

Then, substituting (\ref{q1(phi)}) in (\ref{first_order_problem}) and taking
into account the identities $\cos \varphi =\cos (\pi \tau /2)e$ and $\theta
(\cos \varphi )=(1+e)/2$, gives%
\begin{eqnarray}
\left( \frac{2}{\pi }\right) ^{2}\frac{d^{2}X}{d\tau ^{2}}+X &=&\frac{1}{2}%
A\cos \frac{\pi \tau }{2}\text{,\quad }\frac{dX}{d\tau }|_{\tau =\pm 1}=0
\label{X BVP} \\
\left( \frac{2}{\pi }\right) ^{2}\frac{d^{2}Y}{d\tau ^{2}}+Y &=&\left( \frac{%
1}{2}+2\gamma _{1}\right) A\cos \frac{\pi \tau }{2}\text{,\quad }Y|_{\tau
=\pm 1}=0  \label{Y BVP}
\end{eqnarray}

Recall that boundary conditions in (\ref{X BVP})-(\ref{Y BVP}) eliminate the
singularities of differentiation caused by the formal presence of non-smooth
functions in (\ref{q1(phi)}). It can be verified by inspection that problem (%
\ref{X BVP}) always admits a solution whereas problem (\ref{Y BVP}) is
solvable only under the condition $\gamma _{1}=-1/4$. Note that this number
is identical to that eliminates the resonance term on the right-hand side of
equation (\ref{first_order_problem}). In the present procedure, however, the
number $\gamma _{1}=-1/4$ eliminates the `imaginary' component from
representation (\ref{q1(phi)}) due to the trivial solution of boundary value
problem (\ref{Y BVP}) whereas the `real' component is easily determined from
(\ref{X BVP}). As a result, first-order asymptotic solution (\ref{asymp_sol}%
) takes the closed form%
\begin{eqnarray}
q(t) &=&A\left[ \cos \varphi +\frac{\varepsilon }{8}\left( 2\cos \frac{\pi
\tau }{2}+\pi \tau \sin \frac{\pi \tau }{2}\right) \right] +O(\varepsilon
^{2})  \notag \\
\varphi &=&\omega \left[ 1-\frac{\varepsilon }{4}+O(\varepsilon ^{2})\right]
t\text{,\quad }\tau =\tau \left( \frac{2\varphi }{\pi }\right)
\label{nstt sol}
\end{eqnarray}

Note that solution (\ref{nstt sol}) appears to have the so-called `secular
term' $\tau \sin (\pi \tau /2)$, which is bounded and periodic however with
respect to the original temporal argument $t$.

In order to compare solutions (\ref{PL sol}) and (\ref{nstt sol}), let us
fix the amplitude parameter as $A=1.0$ in (\ref{nstt sol}) and then
determine the corresponding amplitude parameter of solution (\ref{PL sol})
in order to achieve the same initial value $q(0)$. Then, selecting $%
\varepsilon =0.25$ and keeping the three terms of Fourier expansion of order 
$\varepsilon $ as shown in (\ref{PL sol}), gives a very good match of
solutions (\ref{PL sol}) and (\ref{nstt sol}) in terms of the coordinate $q$%
. This is due to quite a rapid convergence of the Fourier series in (\ref{PL
sol}) as seen from its coefficients. However, since differentiation slows
down the convergence then some mismatch in accelerations occurs between the
two solutions as shown in Fig. 13. In the diagram, the dashed line
represents numerical solution produced by the built-in Mathematica%
\registered\ solver.

\subsection{Nonsmooth time decomposition}

In this subsection, the original temporal argument $t$ is replaced by a
sequence of nonsmooth arguments $\{s_{i}\}$ ($i=1,2,...$) running within the
same standard interval $0\leq s_{i}\leq 1$. This brings advantage of using
bounded time arguments and, as shown below, provides a convenient
description of impulsively loaded dynamical systems. As a mathematical
basis, consider the ramp function, 
\begin{equation}
s\left( t;d\right) =\frac{1}{2}\left( d+\left\vert t\right\vert -\left\vert
t-d\right\vert \right)  \label{nstt24}
\end{equation}%
and its first order generalized derivative, $\dot{s}\left( t;d\right) $,
with respect to the temporal argument, $t$; see Figs.14 and 15, respectively.

Such kind of functions are quite common for signal analyses \cite%
{Jackson:1991Addison}. In our case, however, physical interpretation of
these functions represented in Fig. 12 (c) are employed. Namely, the
function $s\left( t,d\right) $ describes positions of a very small perfectly
stiff bead located initially at the origin $x=0$ as shown by the dashed
circle. The initial time instance $t=0$ associates with the event when this
bead is struck on the left by the identical bead with the velocity $v=1$.
Due to the linear momenta exchange, the reference bead starts moving by the
law $x=x(t,d)$ until it hits the third bead located at $x=d=1$. Now, let us
consider an infinite chain of perfectly stiff identical beads on a straight
line at the points $x_{i}$ ($i=0,1,...$). As soon as no energy loss is
assumed, any currently moving bead has the same velocity, $v=1$. As a
result, the linear momentum is translated with the constant speed, whereas
every bead moves only in between its two impact interactions with the
neighbors. The motion of the $i$th bead is described by 
\begin{equation}
s_{i}\left( t\right) =s\left( t-t_{i},d_{i}\right) =\frac{1}{2}\left(
d_{i}+\left\vert t-t_{i}\right\vert -\left\vert t-t_{i+1}\right\vert \right)
\label{nstt25}
\end{equation}%
where $t_{i}$ is the first interaction time of the $i$th bead, and $%
d_{i}=t_{i+1}-t_{i}$ is the time interval between its two consecutive
interactions with its neighbors.

Due to the unit velocity, the variable $s_{i}$ can play the role of `local'
time associated with the bead moving within the interval $x_{i}<x<x_{i+1}$
while the `global' time $t$ runs together with the linear momentum across
the entire chain of beads. However, for any sequence of time instances, $%
\Lambda $ = $\left\{ t_{0},t_{1},...\right\} $, the global time, $t\in
\lbrack t_{0},\infty )$, can be expressed through the sequence of local
times, $\{s_{i}\}$, in the form \cite{Pilipchuk:2002CSF} 
\begin{equation}
t=\sum\limits_{i=0}^{\infty }\left( t_{i}+s_{i}\right) \dot{s}_{i}
\label{nstt26}
\end{equation}%
where the basis $\{\dot{s}_{i}\}$ obeys the following table of products 
\begin{equation}
\text{ }\dot{s}_{i}\dot{s}_{j}=\dot{s}_{i}\delta _{ij}  \label{nstt27}
\end{equation}

As a result, the following `functional linearity' property holds for quite a
general process $x(t)$ 
\begin{equation}
x\left( t\right) =x\left( \sum\limits_{i=0}^{\infty }\left(
t_{i}+s_{i}\right) \dot{s}_{i}\right) =\sum\limits_{i=0}^{\infty }x\left(
t_{i}+s_{i}\right) \dot{s}_{i}  \label{nstt28}
\end{equation}

As mentioned at the beginning of this subsection, the area of applications
of time substitution (\ref{nstt26}) includes impulsively loaded dynamical
systems. In particular, taking into account (\ref{nstt28}) and
differentiation rules \cite{Pilipchuk:2002CSF}, eliminates the external
impulses from dynamical system, 
\begin{eqnarray}
\mathbf{\dot{x}} &=&\mathbf{f}\left( \mathbf{x},t\right) +\sum_{i=0}^{\infty
}\mathbf{p}_{i}\delta \left( t-t_{i}\right) \text{,\qquad }\mathbf{x}\left(
t\right) \in R^{n}  \label{nstt29} \\
\mathbf{x} &\equiv &0\text{,\qquad }t<t_{0}  \notag
\end{eqnarray}%
where $\mathbf{f}\left( \mathbf{x},t\right) $ is regular vector-function and 
$\mathbf{p}_{i}$ are vectors characterizing magnitudes and directions of the
impulses.

The new system is considered then in the local (bounded) time arguments $%
\{s_{i}\}$ by using general analytical tools such as Lie series expansions.
Such approach was applied to the Duffing oscillator with no linear stiffness
under sine modulated random impulses \cite{Pilipchuk:2002CSF}, 
\begin{equation}
\ddot{x}+\zeta \dot{x}+x^{3}=B\sin t\sum_{i=0}^{\infty }\delta \left(
t-t_{i}\right)  \label{nstt30}
\end{equation}%
where $\zeta $ is a constant linear damping coefficient, and $B$ is the
amplitude of modulation.

The distance between any two sequential impulse times is given by 
\begin{equation*}
d_{i}=t_{i+1}-t_{i}=\frac{\pi }{12}\left( 1+\beta \eta _{i}\right)
\end{equation*}%
where $\eta _{i}$ are random real numbers homogeneously distributed on the
interval $\left[ -1,1\right] $, and $\beta $ is a small positive number, $%
0<\beta \ll 1$.

Introducing the state vector $\mathbf{x=}\left( x,\dot{x}\right) ^{T}\equiv
\left( x_{1},x_{2}\right) ^{T}$, brings system (\ref{nstt30}) to the
standard form (\ref{nstt29}), where 
\begin{equation*}
\mathbf{f}\left( \mathbf{x}\right) =\left( 
\begin{array}{l}
x_{2} \\ 
-\zeta x_{2}-x_{1}^{3}%
\end{array}%
\right) \text{,\quad }\mathbf{p}_{i}=\left( 
\begin{array}{l}
0 \\ 
B\sin t_{i}%
\end{array}%
\right)
\end{equation*}

Note that oscillator (\ref{nstt30}) is a modification of the well known
oscillator, $\ddot{x}+\zeta \dot{x}+x^{3}=B\sin t$, considered by Ueda \cite%
{Ueda:1979JSP} as a model of nonlinear inductor in electrical circuits. In
particular, the result of work \cite{Ueda:1979JSP}, as well as many further
investigations of similar models, revealed the existence of stochastic
attractors often illustrated by the Poincare diagrams. Similar diagrams
obtained under irregular snapshots can be qualified as `stroboscopic'
diagrams. The results of the computer simulations \cite{Pilipchuk:2002CSF}
show that some irregularity of the pulse times can be used for the purposes
of a more clear observation of the system orbits in the stroboscopic
diagrams subjected to a chain of significant structural changes as the
parameter $B$ increases. When repeatedly executing the numerical code, under
the same input conditions, such a small disorder in the input results some
times in a less noisy and more organized stroboscopic diagrams. However,
such phenomenon itself was found to be a random event whose `appearance'
depends on the level of pulse randomization as well as the number of
iterations.

\section{Concluding remarks}

This work outlined very different ways to modeling dynamical systems with
discontinuities by choosing proper spatial coordinates or temporal arguments
within the class of nonsmooth functions. Note that nonsmooth transformation
of state vector (\ref{iv7}) can be viewed as a generalization of nonsmooth
coordinate transformation (\ref{oneside q2x}). Whereas transformation (\ref%
{oneside q2x}) is conservative, transformation (\ref{iv7}) possesses built
in energy sinks pumping out some energy from the system whenever it crosses
the image of the barrier in the unfolded space. In contrast to substitution (%
\ref{AP2}), both substitutions (\ref{iv7}) and (\ref{oneside q2x}) are
essentially nonlinear and thus produce essentially nonlinear differential
equations of motion even when out of constraints the equations of motion are
linear. None of the above transformation affects the time variable. In
contrast, the nonsmooth time substitutions described in Section 3
incorporate general temporal symmetries of dynamic processes. Although such
symmetries develop most explicitly through natural motions of elementary
impact models as shown in Fig. 12, the corresponding time substitutions
impose no constrains on class of smoothness of those systems to which such
substitutions are applied. One specific feature of the nonsmooth time
substitutions is that they induce the hyperbolic structure of spatial
coordinates. Further details on the related mathematical properties and
physical interpretations are described in \cite{Pilipchuk:2010Springer}.
Finally note that it is possible to combine different transformations
described in the present survey whenever technical reasons for such
combinations are present; some examples can be found in \cite%
{Pilipchuk:2000JSV} and \cite{Thomsen:2008JSV}.

\pagebreak

\bibliographystyle{plain}
\bibliography{BibAuthors}

\pagebreak

\textbf{FIGURE CAPTIONS}

Fig. 1: Temporal shapes of the force applied to the unit mass $\delta
_{\varepsilon }$ and the corresponding velocity $\theta _{\varepsilon }$ for 
$\varepsilon =0.5$.

Fig. 2: Impulse produced by the `product' of unit-step and Dirac delta
functions versus the model parameter.

Fig.3: Equipotential energy levels in the unfolded configuration plane and a
sample dynamic trajectory obtained under the initial conditions at $t=0$: $%
x_{1}=0.5$, $x_{2}=0.0$, $\dot{x}_{1}=1.0$, and $\dot{x}_{2}=0.0$.

Fig. 4: The beat-wise impact dynamics of the first mass in its original
coordinate.

Fig. 5: The beat-wise impact dynamics of the second mass in its original
coordinate; the energy exchange between the two masses is seen from the
phase shift in their time histories.

Fig. 6: The impact mode trajectories in the unfolded configuration plane:
in-phase and out-of-phase modes (the diagonal lines), and local modes (the
horizontal and vertical lines.)

Fig. 7: A family of phase trajectories of the oscillator with one-sided
barrier at different energy levels for the case $P(q)=q^{2}/2$.

Fig. 8: The phase trajectory of inelastic impact oscillator in the auxiliary
coordinates.

Fig. 9: The original phase plane of the harmonic oscillator with a perfectly
stiff but inelastic one-sided barrier.

Fig. 10: A classic particle reflected by smooth potential barrier.

Fig. 11: The nonsmooth temporal argument $s$ and the particle' motion, $x(t)$%
.

Fig. 12: Three basic impact models generating nonsmooth time substitutions:
(a) positive time, (b) triangular sine-wave time, and (c) nonsmooth time
decomposition.

Fig. 13: Acceleration curves: Poincare-Lindstedt method (thin line),
nonsmooth time substitution (thick line), and numerical (dashed line); $%
\omega =1.0$, and $\varepsilon =0.25$.

Fig. 14: The unit slope ramp function at $d=1.0$.

Fig. 15: First derivative of the ramp function

\begin{figure}
\centerline{\includegraphics[width=1.0\textwidth]{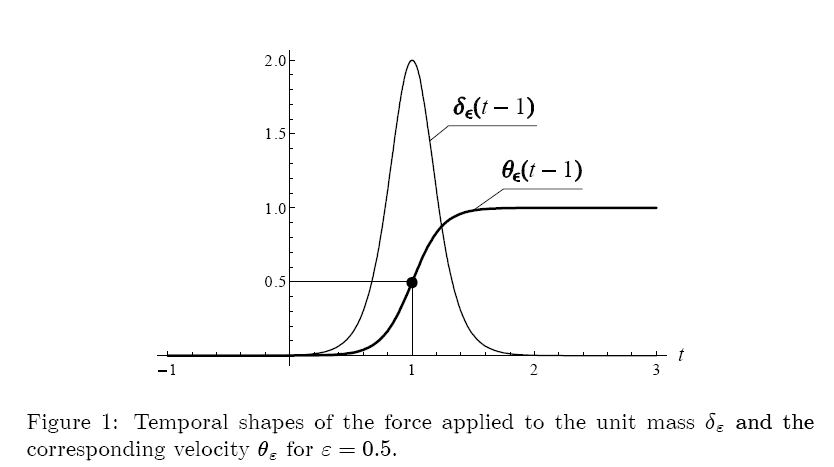}}
\end{figure}

\begin{figure}
\centerline{\includegraphics[width=1.0\textwidth]{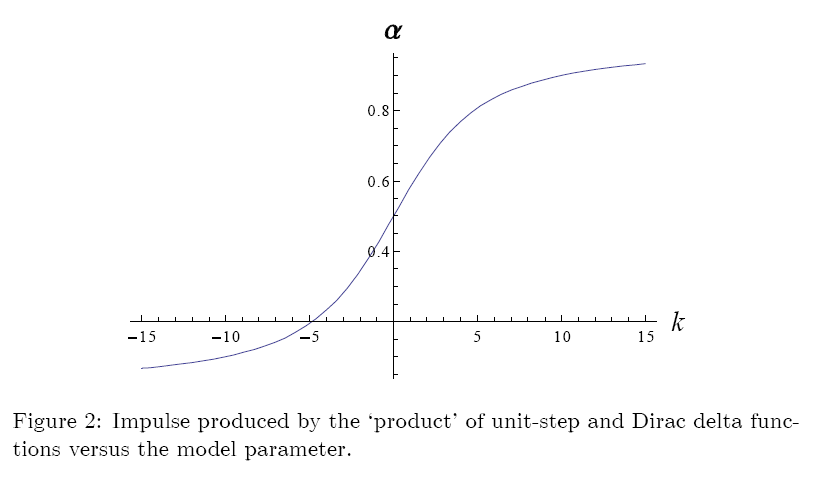}}
\end{figure}

\begin{figure}
\centerline{\includegraphics[width=1.0\textwidth]{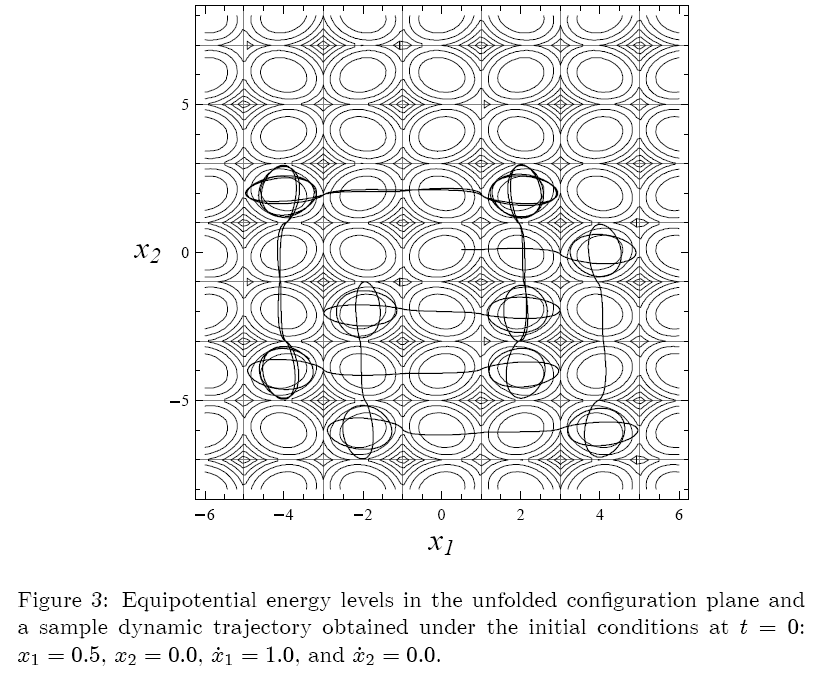}}
\end{figure}

\begin{figure}
\centerline{\includegraphics[width=1.0\textwidth]{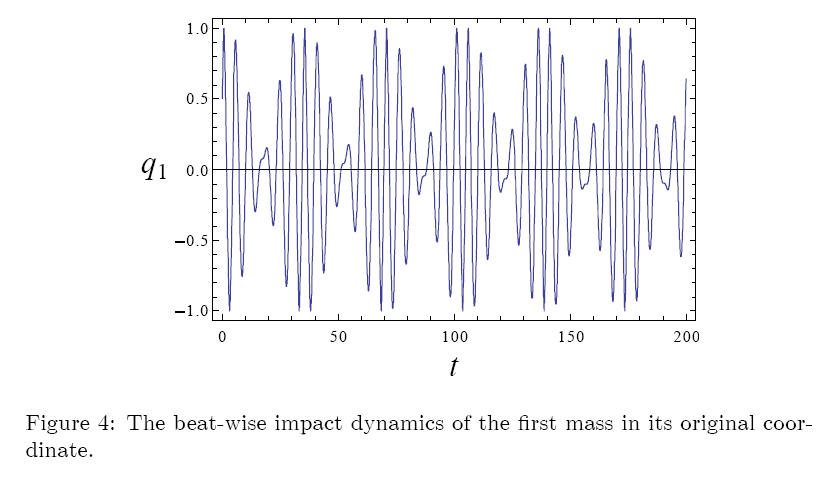}}
\end{figure}

\begin{figure}
\centerline{\includegraphics[width=1.0\textwidth]{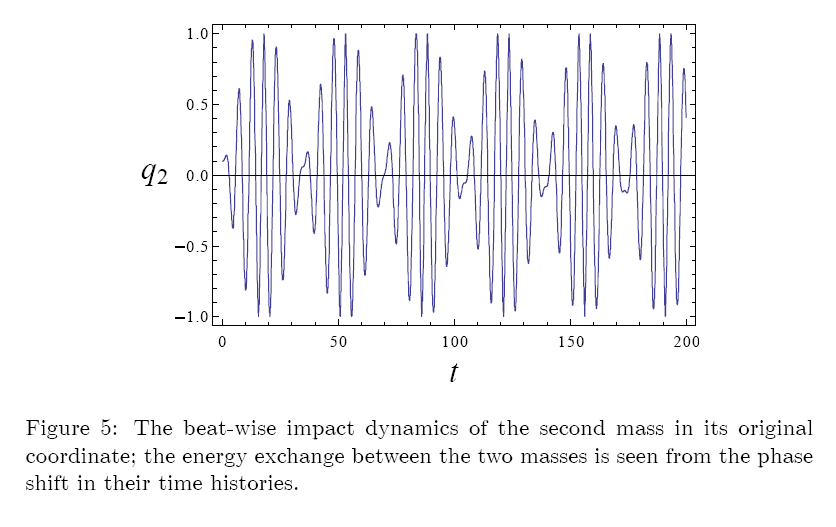}}
\end{figure}

\begin{figure}
\centerline{\includegraphics[width=1.0\textwidth]{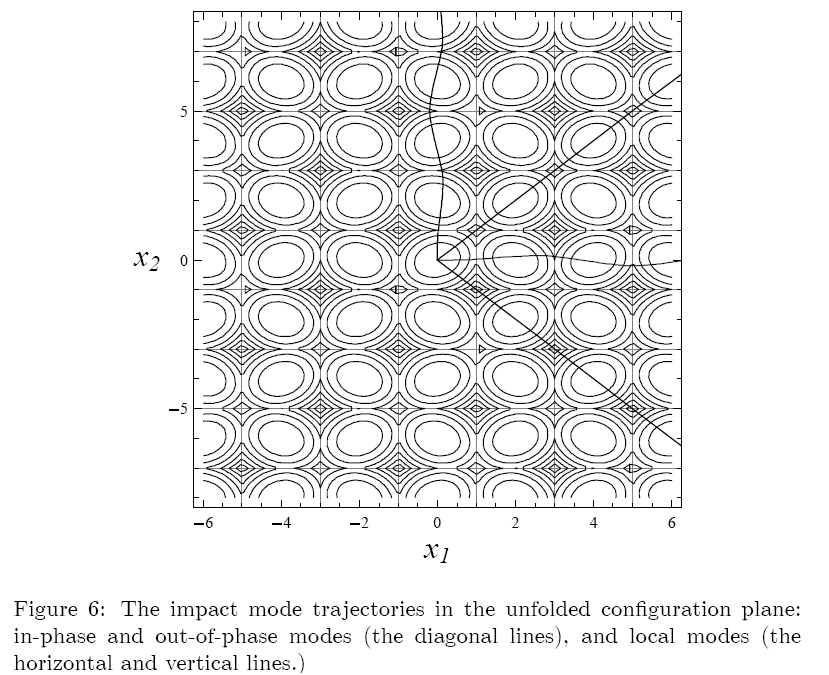}}
\end{figure}

\begin{figure}
\centerline{\includegraphics[width=1.0\textwidth]{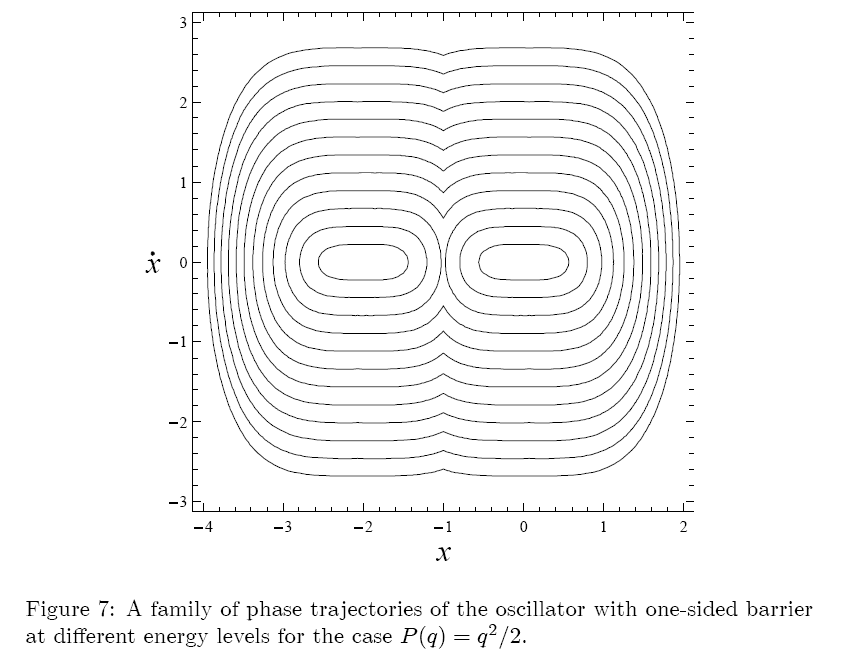}}
\end{figure}

\begin{figure}
\centerline{\includegraphics[width=1.0\textwidth]{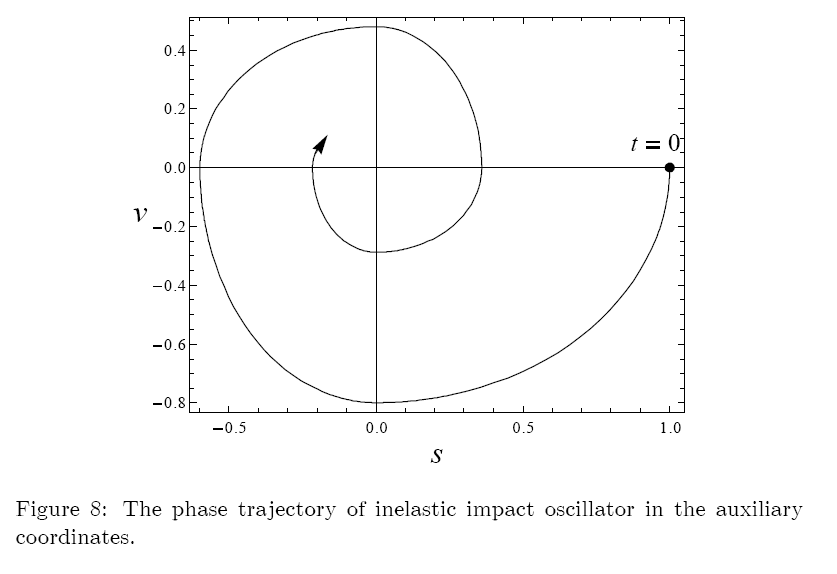}}
\end{figure}

\begin{figure}
\centerline{\includegraphics[width=1.0\textwidth]{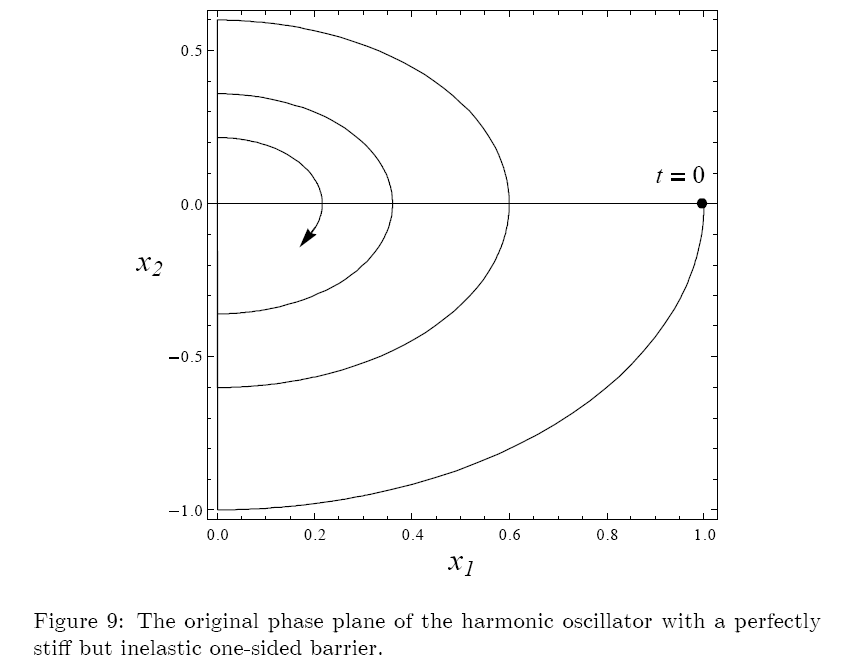}}
\end{figure}

\begin{figure}
\centerline{\includegraphics[width=1.0\textwidth]{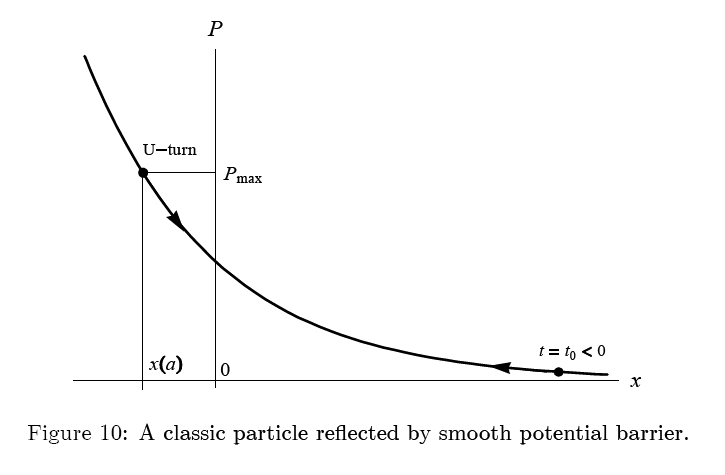}}
\end{figure}

\begin{figure}
\centerline{\includegraphics[width=1.0\textwidth]{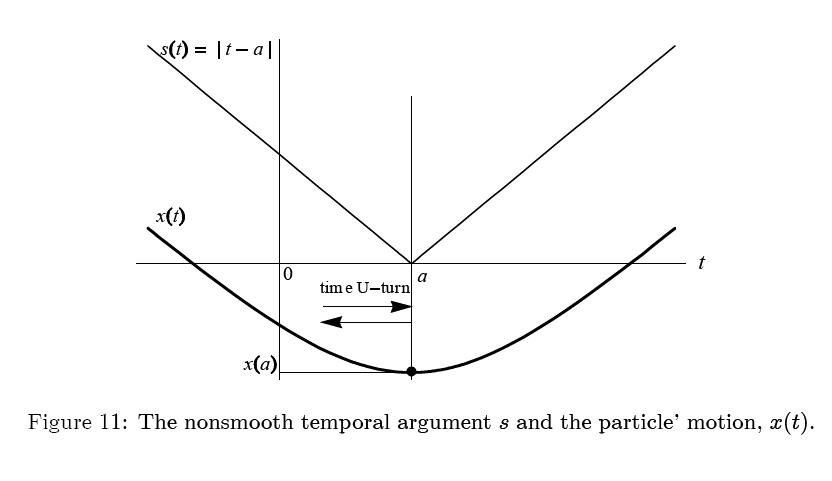}}
\end{figure}

\begin{figure}
\centerline{\includegraphics[width=1.0\textwidth]{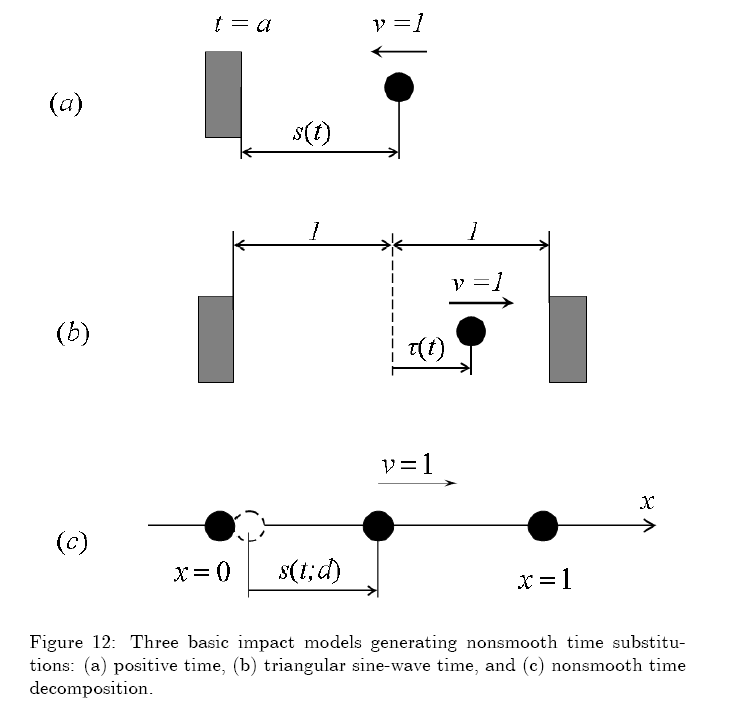}}
\end{figure}

\begin{figure}
\centerline{\includegraphics[width=1.0\textwidth]{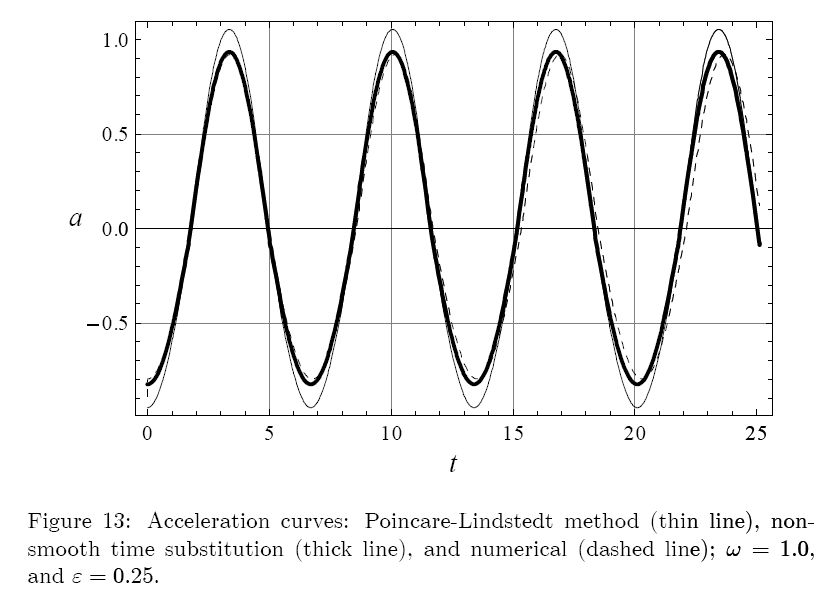}}
\end{figure}

\begin{figure}
\centerline{\includegraphics[width=1.0\textwidth]{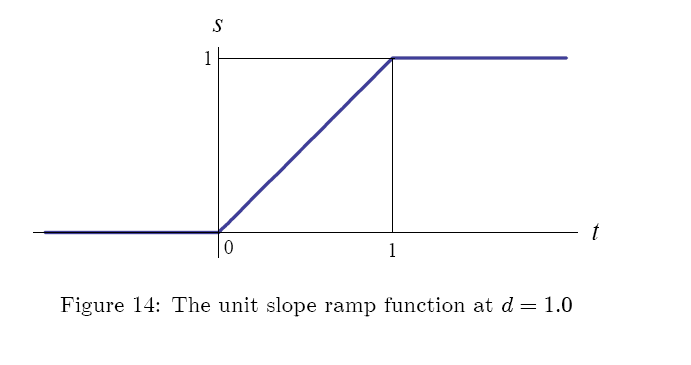}}
\end{figure}

\begin{figure}
\centerline{\includegraphics[width=1.0\textwidth]{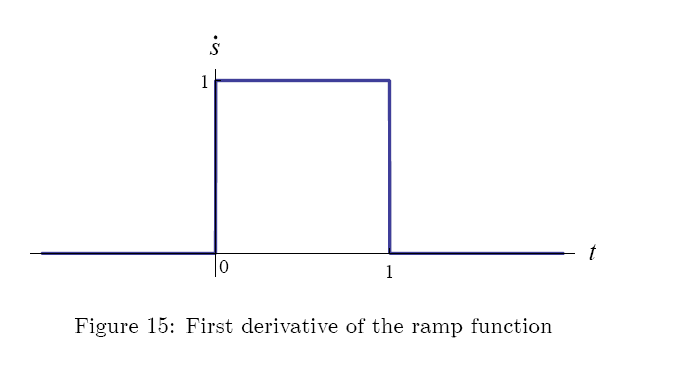}}
\end{figure}

\end{document}